\documentclass[epj,referree]{svjour}
\usepackage{graphics}
\begin{document}
\title{Microwave transport approach to the coherence of interchain hopping %
in (TMTSF)$_2$PF$_6$}
\author{P. Fertey\inst{1}\thanks{\emph{Present address:} Laboratoire de %
Cristallographie et Mod\'elisation des Mat\'eriaux Min\'eraux et Biologiques, %
Universit\'e Henri Poincar\'e, Nancy I,%
  B.P. 239, 54 506 Vandoeuvre-les-Nancy Cedex, France.},M. Poirier\inst{1} %
  \and P. Batail\inst{2}
}                    
\institute{Centre de Recherche en Physique du Solide, D\'ep. de
Physique, Univ. de Sherbrooke, Sherbrooke, Qu\'ebec, Canada J1K 2R1.
\and
Institut National des Mat\'eriaux de Nantes, 2 Rue de la Houssini\'ere, %
 B.P. 32229, %
44 322 Nantes C\'edex 3, France.}

\date{Received: date / Revised version: date}
%
\abstract{
We report a microwave study of the longitudinal and transverse transport
properties of the
quasi-one-dimensional organic conductor (TMTSF)$_2$PF$_6$ in its normal
phase. The
contactless technique have provided a direct measurement of the
temperature profile of the resistivity along the {\bf b'} direction and in
magnetic fields up to 14 T. A characteristic energy scale ($T_x \sim 40 K$) has
been observed which delimits a transient regime from an insulating to a metallic 
behavior. This
anomalous profile is discussed in terms  of the  onset of  coherent
transport properties along  the {\bf b'} direction below 40 K. This is
also supported by the
observation of a finite longitudinal and transverse magnetoresitances
only below 40 K, indicative of a two-dimensional regime. Below $T_x$, 
however, strong deviations
with respect to a Fermi liquid behavior are evidenced.
\PACS{
	  {67.55.Hc}{Transport properties} \and
      {71.27-a}{Strongly correlated electron systems, heavy fermions} \and
      {71.10.Pm}{Fermions in reduced dimensions (anyons, composite fermions, %
      Luttinger liquid, etc.} \and
      {74.70.Kn}{Organic superconductors.}
     } 
} 
\authorrunning{Fertey et al.}
\titlerunning{Transverse coherent transport in (TMTSF)$_2$PF$_6$}
\maketitle
\section{Introduction}
\label{intro}
Due to a
pronounced chain structure, the (TMTSF)$_2$X, [X = PF$_6$, AsF$_6$, ClO$_4$...] Bechgaard salts
have become the prototypical
examples of
quasi-one-dimensional (Q1D) conductors with the highest conductivity
direction parallel to the stacking axis ({\bf a}) of the TMTSF
molecules. Their low temperature properties have attracted much attention 
since various transitions such as incommensurate
spin-density-wave (SDW),
superconductivity, field-induced SDW, quantum Hall effect etc., have been
observed \cite{Ishiguro90}. Recently, the normal phase (i.e. above the
transition temperature
 of the broken symmetry ground states) has attracted much interest. Since the tunneling integral along
the chain direction ($t_a \sim 0.25~eV$) is at least one order of
magnitude larger than the transfer integrals
$t_b$ and $t_c$ in the transverse directions ($t_b \sim 200 $K and $t_c
\sim 10 $K), the organic metallic chains are usually
considered as weakly coupled. Although the coupling along {\bf
c} is likely irrelevant over a large temperature domain, the effective value
of $t_b$ has to be considered to precise the dimensionality of the electron gas in
the normal phase. At sufficiently high temperatures however, the physical properties
are expected to be essentially governed by 1D phenomena.\\
It is well known that, in a strictly 1D interacting electron gas, the Fermi liquid (FL)
picture breaks down and must be replaced by the so-called Luttinger liquid
(LL) description \cite{Schulz91}. Since non-negligible interchain
coupling along the {\bf b'} direction exist in the Bechgaard's salts, 
departure from the LL model \cite{Boies95} migth be induced. Indeed, when
the temperature is progressively lowered, transverse {\bf b'} interactions are expected
to become more effective so that a crossover from a Q1D to a two-
dimensional (2D) electron gas picture should occur: the FL behavior
might be recovered provided that the Coulomb
interactions are not too strong. However, the actual value of the crossover
temperature $T_x$ is highly debated. According to simple
band calculations, the dimensional crossover for the single
particle motion is then expected to occur at $T_x\sim 150-200 K$.This is
in agreement with the temperature dependence of the longitudinal DC
resistivity which is showing a transition regime from a roughly linear behavior
to a $T^2$ profile (indicative of a FL behavior dominated by
electron-electron scattering effects) over that temperature
range \cite{Jerome94}. However, photoemission spectra 
\cite{Dardel93,Zwick97} are incompatible with a FL picture over that
temperature range and early optical experiments with the
light polarized along the transverse {\bf b'} direction
failed to evidence a coherent plasma edge above 50K \cite{Jacobsen83}.
Moreover, deviations from the FL picture have also been observed down to
50 K in NMR experiments, suggesting then an upper bound
value for the crossover temperature \cite{Bourbonnais93,Wzietek93}.
Furthermore, the frequency dependence of the conductivity is well known to
display unusual features \cite{Schwartz98,Vescoli98}: for frequencies above
the effective interchain transfer integral, the electrodynamics is consistent with the
prediction of the LL picture, while at low frequencies, pronounced deviations
with respect to the Drude picture are present \cite{Timusk96}.\\
Among all the experimental approachs used to study the
dimensionality of the electron gas, transverse transport measurements
are particularly relevent to directly probe the interchain
couplings. It was further realized that since the transverse transfer integrals
are small, an electric field applied along the {\bf b'} or {\bf c*}
directions could also act as a probe of the physical properties in the
plane perpendicular to that direction. Early resistivity measurements 
\cite{Jacobsen81b} along the hard axis ({\bf c*}) have shown a non
monotonic behavior of the temperature profile: a maximum of $\rho_c$ was
observed near 80 K. More recently, a strong pressure dependence of this unusual feature was
evidenced \cite{Moser98} and a typical 1D power law profile was found above the
characteristic $\rho_c$ maximum. This maximum was then
ascribed to a broad crossover regime indicative of a deconfinement of
the carriers from the chain axis; this results in a gradual onset of
coherent transport along {\bf b'} below 80 K, suggesting then a FL
behavior in the {\bf a-b'} plane. However, the anisotropy ratio $\rho_c /
\rho_a$ was not found temperature independent as expected from FL
arguments and an incipient Fermi liquid was therefore invoked. These
observations contrast with the work of Gor'kov et
al. \cite{Gorkov98,Gorkov96,Gorkov95} who recently argued that the
longitudinal transport properties, below 60 K
and down to the SDW transition temperature, can be well accounted for in
terms of a weakly interacting Fermi liquid. However, such a
quasi-particle like signature (if ever) should also be detected in the
{\bf b'} transverse direction.\\
Reliable measurements of the transverse transport properties along
{\bf b'} are highly needed to clarify the present controversy.
Unfortunately, since the Bechgaard salts have a pronounced
needle shape whose axis is parallel to the chains, transverse transport
along {\bf b'} is particularly difficult to perform with usual DC
methods. Owing to non-uniform current distributions
between contacts, parasitic contributions from other directions are
likely introduced. These problems can be avoided by using a contactless microwave
technique which allows a better control on the orientation of the
current lines in these organic needles.  In this paper, we report
microwave resistivity data obtained along the transverse directions in
(TMTSF)$_2$PF$_6$ crystals.  These data clearly indicates the
temperature range over which the coherence of interchain hopping sets
in and they confirm strong deviations from a Fermi liquid description.
\section{Experiment}

High quality single crystals of (TMTSF)$_2$PF$_6$ have been
synthesized by the standard electrochemical method with typical dimensions 
($\sim 8\times0.25\times0.
1$~mm$^3$) along the {\bf a}, {\bf b'} and {\bf c*} axes respectively. 
Such a needle geometry is not suitable for precise
measurement of the transverse transport properties; this is particularly
true for our microwave technique which yields very accurate data only
when the electric field is oriented along the needle's axis. Each needle
was therefore cut into three pieces in order to perform the measurements
along the {\bf a}, {\bf b'} and {\bf c*} axes (natural faces of the
single crystals) on the same single crystal.  Each piece was then cut in
small blocks so that it could be reconstructed with the shape of a needle
having one of the crystal directions as its axis. We used
a conventional cavity perturbation technique at 16.5 GHz \cite{Fertey97}
to obtain the electrical resistivity along each crystal axes as a
function of the temperature (2-300 K) and a magnetic field (0-14 Tesla)
applied along the {\bf c*} direction. Unfortunately, due to the
microwave resonator design, no data could be collected with both the
magnetic field and the electric field along the {\bf c*} axis.  To
prevent microcraks the samples were slowly cooled to the lowest
temperature at 0.6K/min. The temperature was monitored either with a Si
diode (zero field) or a capacitor sensor (field up to 14T).
\section{Results and discussion}

\begin{figure}[tp]
\centerline{\epsfbox{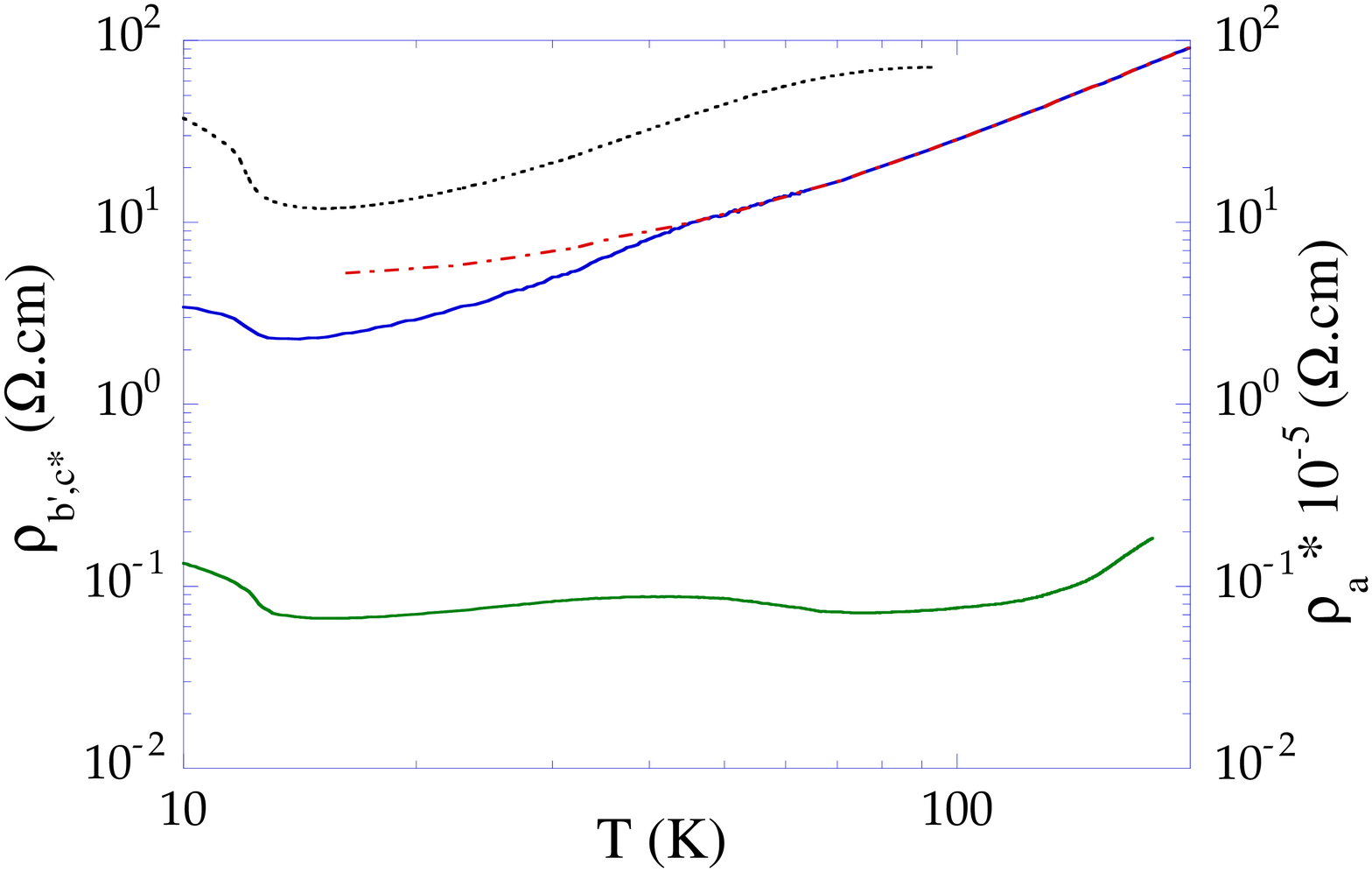}}
\caption{%
Microwave resistivity of (TMTSF)$_2$PF$_6$ as a function of temperature
along {\bf a}
(full), {\bf b'} (dashed) and {\bf c*} (dotted) axes.  The dashed-dotted
line mimics the behavior of the dc resistivity along the chain axis.}
\label{fig:1}
\end{figure}

Along the high conductivity axis, the microwave resistivity has
been determined by using the Hagen-Rubens limit (skin depth
regime). Since the conductivity is much lower along the
{\bf b'} and {\bf c*} axes, the microwave data were rather analyzed in
the framework of the metallic limit of the quasistatic
approximation \cite{Musfeldt95}. We report in Figure \ref{fig:1}, the
temperature profile of the resistivity along the ({\bf a}), ({\bf b'})
and ({\bf c*}) axes of (TMTSF)$_2$PF$_6$ in the normal phase for zero
magnetic field value. The orders of magnitude are in good agreement with
published DC results \cite{Jacobsen81b,Moser98} and
the SDW phase is evidenced for each crystal direction by an abrupt
resistivity increase below 12 K.

\input epsf.tex
\epsfxsize 6.5 cm

\begin{figure}[tp]
\centerline{\epsfbox{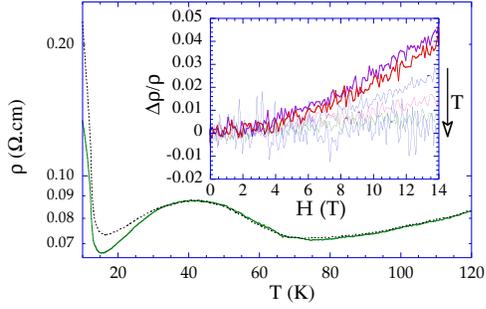}}
\caption{%
Microwave resistivity along {\bf b'}, in zero (full) and 14 Tesla (dotted) 
magnetic field
values. Inset: relative magnetoresistivity measured at 16, 18, 20, 25,
35 and 80 K. The  right arrow indicates the temperature variation.}
\label{fig:2}
\end{figure}

\begin{figure}[tp]
\centerline{\epsfbox{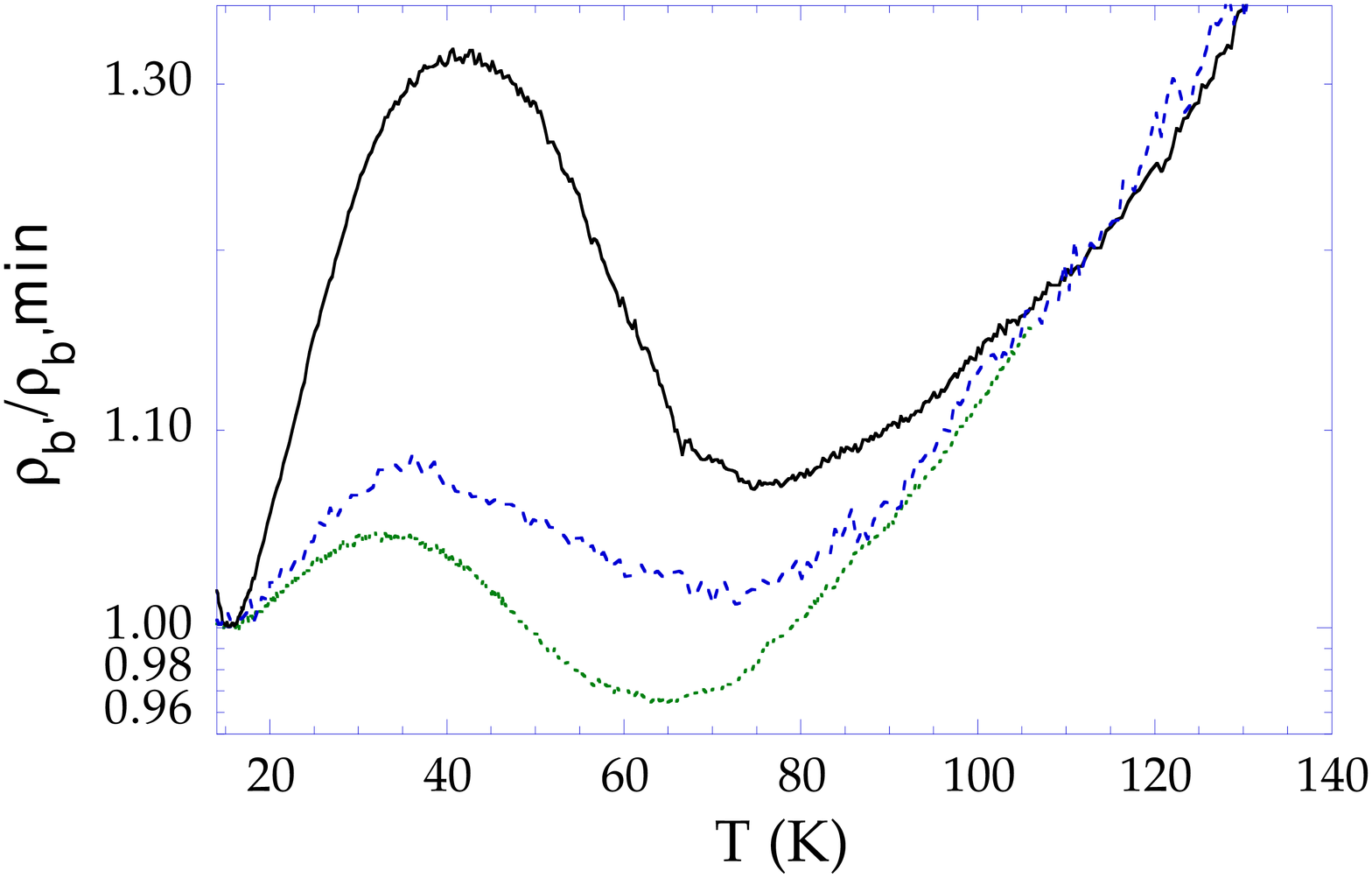}}
\caption{%
Sample effect on $\rho_{b'}$ at intermediate
temperatures for sample 1 (full), 2 (dashed) and  3 (dotted).}
\label{fig:3}
\end{figure}

Along the chain axis ({\bf a}), the usual metallic behavior is observed
down to 13.6 K, where the resistivity reaches a minimum.  Interestingly,
the microwave resistivity profile displays clearly, near 45 K, a change
of slope, when the DC one usually shows a single quadratic behavior
below 100 K (dashed-dotted line in Figure \ref{fig:1}). Along the least conducting
direction ({\bf c*}), the microwave resistivity profile is consistent
with the DC curve \cite{Moser98}: it increases first when the
temperature is decreased from 300 K (not shown on the figure), reaches a
maximum near 90 K and recovers a metallic behavior on further cooling.
The minimum value is obtained near 15.3 K. The {\bf b'} resistivity
presents definitely a different profile, being almost flat (below 120 K)
on the logarithmic scale compared to the other crystal directions. This
profile is shown in more details in Figure \ref{fig:2}.  On lowering the
temperature, the resistivity $\rho_{b'}$ first decreases monotonically
down to a local minimum around 75 K, increases slightly to reach a local
maximum near 40 K and decreases again down to 15 K before entering the SDW
phase below 12 K.  This peculiar profile observed between 12 and 80 K is
weakly sample dependent: this is illustrated in Figure \ref{fig:3} where we
compare the normalized resistivity (relative to the value just above the
SDW transition) obtained on three different samples
(different batches).  Such a dependency can be explained by two factors:
i) a sligth misalignment relative to one another of the small crystals
used in the needle's construction; ii) a different impurity content in
crystals of different batches.  A correlation with the latter factor is
difficult to evaluate for the moment.  However, it seems clear from
Figure \ref{fig:3} that the local maximum of $\rho_{b'}$ around 40 K is intrincic to
this crystal direction.  Its absence on the DC resistivity
profile \cite{Jacobsen81b} could signify that the latter is significantly
polluted by different components of the resistivity tensor as previously
mentioned.

\input epsf.tex
\epsfxsize 6.5 cm

\input epsf.tex
\epsfxsize 6.5 cm

The emergence of an insulating behavior below 70 K ($d\rho_{b'}/dT <
0$ in Figure \ref{fig:2}) refutes the possible existence of quasi-particle states down to 40 K.
A Fermi liquid description of interacting electrons above 40 K would,
indeed, have required a quadratic temperature profile of both the $\rho_a$
and $\rho_{b'}$ components. Between 50 K and 70 K, $\rho_{b'}$ is better
understood by assuming a Luttinger liquid behavior along the stacks: this yields a power-law
increase $\rho_{b'}(T) \sim T^{-2\alpha}$, where $\alpha$ is the exponent of
the single-particle density of states of the LL \cite{Moser98}. However, due to the reduced
temperature domain used to
fit the power law and the slight sample dependency of
$\rho_{b'}$ in this temperature range, the exponent $\alpha$ might be only
approximative ($2\alpha\sim 0.3$) and prevent in turn reliable
determination value of the exponent $K_\rho$ characterizing the charge degrees of
freedom of a LL.

\begin{figure}[tp]
\centerline{\epsfbox{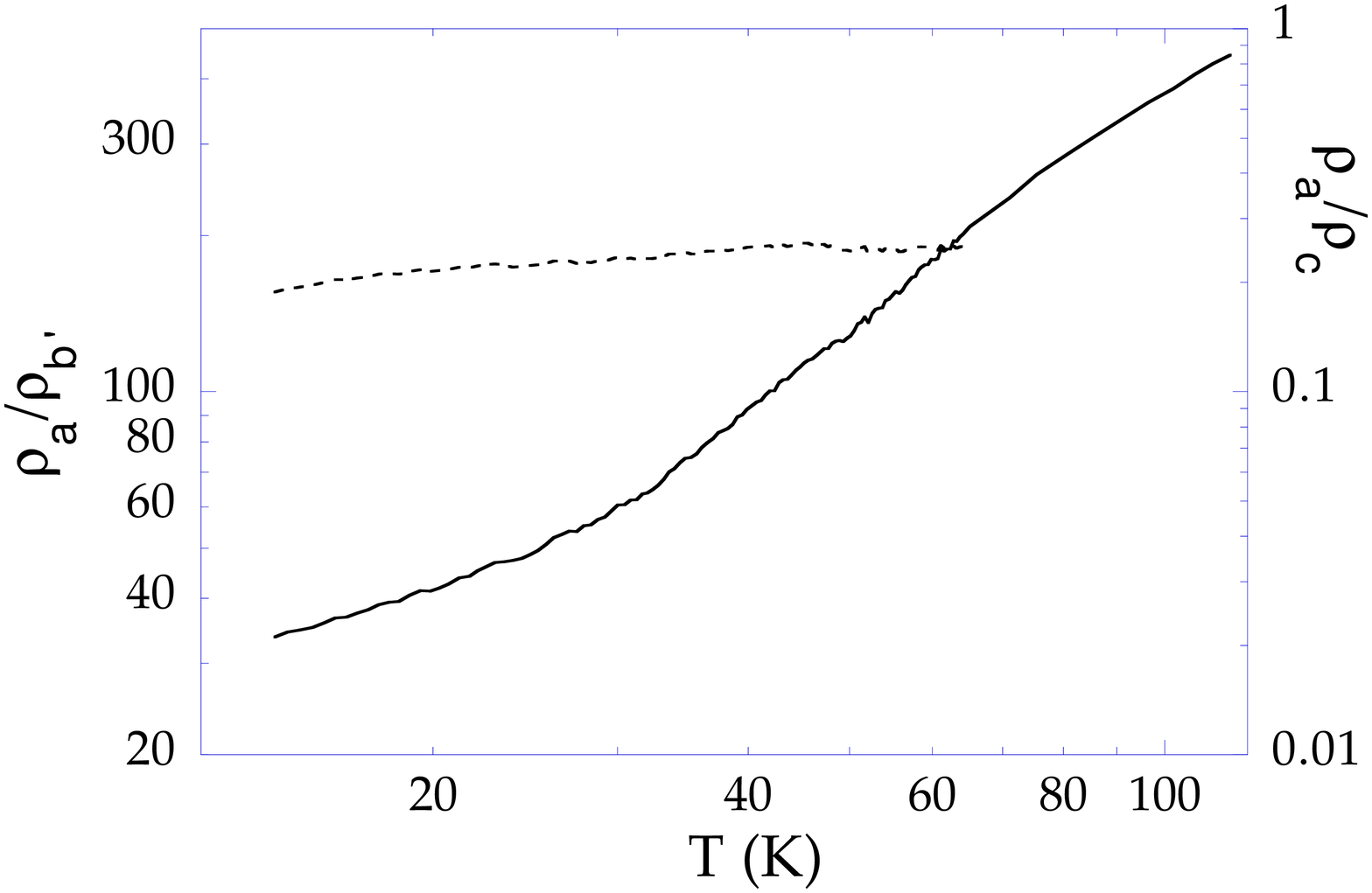}}
\caption{%
Temperature profiles of the anisotropy ratios $\rho_{{\bf a}}/\rho_{{\bf
b'}}$ (line) and $\rho_{{\bf a}}/\rho_{{\bf
c*}}$ (dashed).}
\label{fig:4}
\end{figure}

The resistivity maximum observed around 40 K mimics the temperature profile of $\rho_c$ which
also displays a resistivity maximum at higher temperature, near 80
K \cite{Jacobsen81b,Moser98}.  Therefore, it could be attributed to the onset of a deconfinement
of the carriers and the restauration of a 2D conductivity regime in the {\bf a-b}
plane below 40 K. This is supported by the observation of an important increase of
the resistivity when a magnetic field is applied perpendicularly to the
2D plane of motion (Figure \ref{fig:2}), only for temperatures below 40 K. 
This is exemplified in the inset of
Figure \ref{fig:2} which displays the variation of the resistivity
relatively to its zero field value as a function of the magnetic field
for temperatures between 16 and 80 K. It is well known that in the
Bechgaard salts, a magnetic field applied along the hard axis {\bf c*}
confines the motion of the carriers to the chain axis.  Therefore, a
reduction of the metallic behavior parallel to {\bf b'}  is expected
if there is some coherent motion  along that direction. Below 40 K, the progressively 
increasing magnetoresistivity
can then be interpreted as the signature of the 2D motion. On the contrary, since no 
magnetoresistivity is
observed above 40 K, transverse coherent
hopping is apparently absent and coherent motion is confined to the
organic stacks. This
observation of a dimensional crossover around 40 K supports the
predictions of the renormalization group theory: a Luttinger liquid
picture may persist down to the low temperature region when many-body
effects on interchain hopping are considered \cite{Bourbonnais85}.
Indeed, one-dimensional many-body effects are expected to lower the efficiency
of interchain tunneling, thereby decreasing its amplitude. However, transients
effects are likely associated to such a crossover since the
deconfinement region is not sharply defined but spreaded out in
temperature: in the coherent regime (16-30 K), $\rho_{b'}$ does not show
a quadratic temperature profile (the power law exponent ranges from 0.37 to 0.075
for the three samples studied). This signifies that important
deviations from a FL quasi-particle transport exist along the {\bf b'}
axis, as exemplified by the temperature profile of the microwave
resisitivity anisotropy ratio $\rho_{{\bf a}}/\rho_{{\bf b'}}$ and
$\rho_{{\bf a}}/\rho_{{\bf c*}}$ presented in Figure \ref{fig:4}.
\input epsf.tex
\epsfxsize 6.5 cm
Below 70 K, $\rho_{{\bf
a}}/\rho_{{\bf c*}}$ is practically constant: it
can be reasonably fitted by a $\sim T^{0.15}$ law in clear contrast with the results
of Moser et al. \cite{Moser98}, ($\sim T^{0.5}$) over the same temperature
range. On the contrary, $\rho_{{\bf a}}/\rho_{{\bf
b'}}$ continuously decreases from 300 K down to the SDW
transition temperature. Only a change of curvature is observed around the
dimensional crossover temperature near 40 K, seemingly in
contradiction with a constant ratio expected for a true FL behavior.

\input epsf.tex
\epsfxsize 6.5 cm
\begin{figure}[tp]
\centerline{\epsfbox{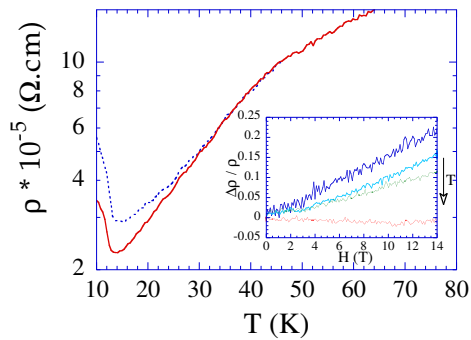}}
\caption{%
Temperature profile of the chain axis microwave resistivity in zero
(full) and 14 Tesla (dotted) magnetic field values. Inset: relative
magnetoresistance measured at 14, 16, 18 and 80 K. The  right arrow
indicates the temperature variation.}
\label{fig:5}
\end{figure}

The onset of coherence transport along the transverse direction is
also supported by some features observed on the microwave resistivity
along the chain axis. We show, in Figure \ref{fig:5}, a significant increase of
the resisticivty for a magnetic field of 14 Tesla applied along {\bf
c*}.  As seen in the inset, a significant magnetoresistance is again
observed only for temperatures below 40 K, consistently with a
dimensional crossover over that range. The change of slope precedingly
identified on $\rho_{a}$ around 45 K in zero field thus signals the
onset of transverse coherent transport.

\section{Conclusion}

The microwave resistivity data reported in this paper for
(TMTSF)$_2$PF$_6$ crystals clearly show the onset of coherent 
transport properties along the intermediate
conductivity direction {\bf b'}. This
temperature scale is evidenced by a resistivity maximum around 40 K 
along the {\bf b'} axis and supported by a progressively
increasing magnetoresistance below 40 K, when a magnetic field is applied along
{\bf c*}. Similar effects observed on the temperature profile of the
longitudinal resistivity $\rho_{a}$ confirm the dimensional crossover.
Furthermore, a temperature profile analysis of
$\rho_{b'}$ has failed to detect any clear-cut Fermi liquid component 
in the whole normal phase domain. Our results seems to be in  good agreement with previous
results such as NMR \cite{Bourbonnais93,Wzietek93}, photoemission \cite{Dardel93}
or optical data \cite{Timusk96}, which claim  the abscence of any quasi-particle
features. These results, however, contrast with
reflectance \cite{Schwartz98,Vescoli98} or very low temperature
magnetotransport \cite{Danner94} data, which seem to be well
described by bare band parameters. A detailed theoretical
framework that would clarify why some
experimental probes
are apparently strongly sensisitve to many-body effect while others do not,
is missing so far.
\begin{acknowledgement}
The authors are grateful to J. Beerens for giving access to his 14
Tesla experimental set-up, to C. Bourbonnais for critical reading of the
manuscript and useful suggestions during the course of this work,
M. Castonguay and
J. Corbin for technical assistance.  This work was supported by grants
from
the Fonds
pour la Formation de Chercheurs et l'Aide
\`a la Recherche of the Government of Qu\'ebec (FCAR) and from the Natural
Science
and Engineering Research Council of Canada (NSERC).
\end{acknowledgement}

\end{document}